\input{aipcheck}

\documentclass[
   ,final            % use final for the camera ready runs
  ]
  {aipproc}

\layoutstyle{6x9}
\renewcommand\XFMtitleblock{%
\XFMtitle
\let\XFMoldpar\par
\def\par{\XFMoldpar\def\par{\space
    for the VERITAS Collaboration\footnote{\protect\url{http://veritas.sao.arizona.edu/}}\XFMoldpar}}%
 \XFMauthors
 \let\par\XFMoldpar
 \XFMaddresses
 \XFMabstract
 \vspace{5pt}%
 \XFMkeywords
 \XFMclassification
}
\usepackage{wrapfig}
\usepackage{multicol}
\usepackage{sidecap}
\begin{document}

\title{VERITAS Observations of the Crab Pulsar}


\author{Benjamin Zitzer}{
  address={Argonne National Laboratory, 9700 S. Cass Ave. Lemont, IL, USA}
}
%
%\author{the VERITAS Collaboration}{
%  address={\url{http://veritas.sao.arizona.edu}}
%}
\begin{abstract}
The Crab pulsar has been widely studied across the electromagnetic spectrum from radio to gamma-ray energies. The exact nature of the emission processes taking place in the pulsar is a matter of broad debate. Above a few GeV the energy spectrum turns over suddenly. The shape of this cutoff can provide unique insight in to the particle acceleration processes taking place in the pulsar magnetosphere. Here we discuss the detection of pulsed gamma-rays from the Crab Pulsar above 100 GeV with the VERITAS telescopes in the context of measurements made with the Fermi space telescope below 10 GeV. Limits on the level of flux enhancement of emission correlated with giant radio pulses and dispersion due to Lorentz invariance violation effects will also be presented. 
\end{abstract}

\maketitle

%%%%%%%%%%%%%%%%%%%%%%%%%%%%%%%%%%%%%%%%%%%%
%% MAINMATTER
%%%%%%%%%%%%%%%%%%%%%%%%%%%%%%%%%%%%%%%%%%%%

\section{INTRODUCTION}

 The Crab nebula is believed to be a remnant of a supernova observed in 1054 A.D. Much later, a high-$\dot{E}$ pulsar with a period of $\sim$33ms was observed in the system \cite{Crab1968}. 

All 117 known $\gamma$-ray pulsars, including the Crab, show a spectral cutoff above a few GeV \cite{Celik2012}. Traditional pulsar models attribute this cutoff to curvature radiation within the magnetosphere. Measuring the spectral break energy and shape of the cutoff helps to contrain these models. MAGIC's observations of of the Crab pulsar revealed significant pulsations at 60 GeV with hints of pulsar signals at energies higher than 120 GeV \cite{MagicSci2008}, giving the first view of the possiblilty of a non-exponential cutoff spectrum of a pulsar. Pulsed emission was later detected by VERITAS above 100 GeV, rejecting the exponential cutoff model at the 5.6$\sigma$ level \cite{OtteSci2011}.    

%\section{Data Selection and Timing Analysis}
\section{DATA SELECTION AND TIMING ANALYSIS}

VERITAS (Very Energetic Radiation Imaging Telescope Array System) is an array of four IACTs (Imaging Atmospheric Cherenkov Telescopes) located south of Tucson, Arizona, USA \cite{Galante2012}. VERITAS collected 107 hours of low zenith observations on the Crab from the start of four telescope operations in 2007 to 2011. Data quality selection requires a clear atmospheric conditions, based on infrared sky temperature measurments and nominal hardware operation. Event selection that was applied to the data was optimized a priori by assuming a power-law spectrum with an index of -4.0 and a normalizaion of a few percent of the Crab at 100 GeV \cite{OtteSci2011}. Data reduction followed the standard methods, yeilding consistant results with two analysis packages \cite{Analysis}. 

The Jordell Bank timing ephemeris was used to obtain the timing parameters for the pulsar analysis \cite{JordellBank}. Barycentering was done with two custom codes and tempo2 \cite{Tempo2}. Applying the H-test  \cite{DeJager1994} to this data set yields a H vaule of 50, giving a 6.0$\sigma$ significance \cite{OtteSci2011}. Using a Li \& Ma significance \cite{LiMa1983} using defined ON and OFF windows gives a 8.8$\sigma$ signficance \cite{OtteSci2011}. An unbinned maximum liklihood fit determined the positions of P1 and P2 to be -0.0023$\pm$0.0020 and 0.0398$\pm$0.002, respectively. The ratio of pulsed events in P2 over P1 is 2.4$\pm$0.6 \cite{OtteSci2011}.  An additional 20 hours of data was taken during the 2011-2012 observing season. The combined Li \& Ma test signifance is 10.7$\sigma$. The pulse profile of the entire data set is shown in figure 1.

\begin{figure}[t]
%\begin{wrapfigure}{l}{0.5\textwidth}
\centering
  \includegraphics[height=.3\textheight]{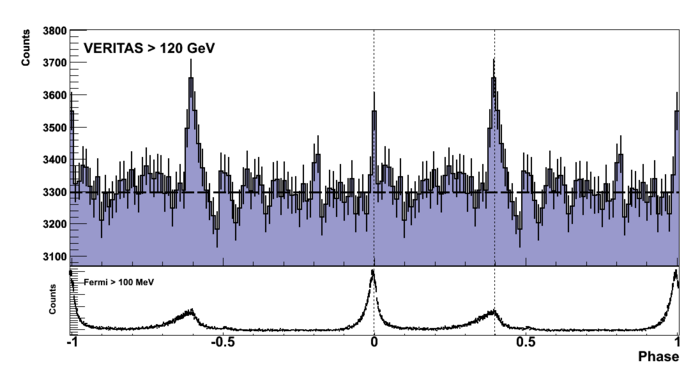}
  \caption{Combined VERITAS Crab pulsar phase histogram for all quality data between 2007 and 2012. The Fermi-LAT phase histogram is also shown (below). }			
%\end{wrapfigure}
\end{figure}

%\section{Spectra and Interpretation}
\section{SPECTRA AND INTERPRETATION}

The combined pulsed excess of P1 and P2 of the 2007 to 2011 data set is fit to a powerlaw function: $dN/dE = N_{0}(E/150 GeV)^{-\Gamma}$, with the flux normalization, $N_{0}=(4.6\pm0.6_{stat}+2.4_{sys}-1.4_{sys})$x10$^{-11}$TeV$^{-1}$cm$^{-2}$s$^{-1}$, and the spectral index, $\Gamma=3.8\pm0.5_{stat}\pm0.2_{sys}$. The analysis energy threshold is 120 GeV \cite{OtteSci2011}. 

The combined Fermi-LAT and VERITAS spectral energy distribution is shown figure 2. The broken powerlaw fit of the data yeilds a $\chi^2$ fit of 13.5 for 15 degrees of freedom. This provides a better fit than the exponential cut-off fit, yeilding a $\chi^2$ value of 66.8 for 16 degrees of freedom. The fit probability of 3.6x10$^{-8}$ derived from the $\chi^{2}$ value excludes the exponential cut-off as a viable parameterization of the Crab pulsar spectrum \cite{OtteSci2011}. These findings were later confirmed by the MAGIC collaboration \cite{MagicCrab}.

The detection of pulsed $\gamma$-ray emission above 100 GeV provides strong contraints on the $\gamma$-ray radiation mechanisms and location of the acceleration regions. The shape of the spectrum above the break cannot be attributed to curvature radiation since that would require an exponentially shaped cut-off and a radius of curvature larger than the light cyclinder radius. It is therefore likely that $\gamma$-ray production is dominated by an emission mechianism other than curvature radiation, or a second emission process becomes dominate above the spectral break energy.

%\begin{wrapfigure}{L}{0.5\textwidth}
\begin{figure}
\centering
\includegraphics[height=.35\textheight]{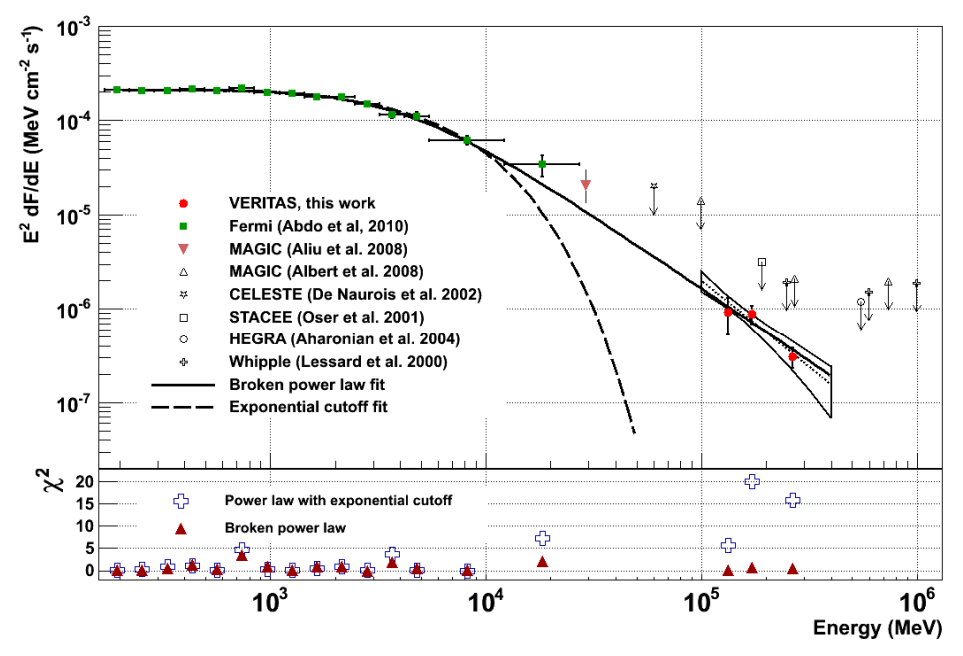} 

  \caption{Spectral Energy Distribution of the combined VERITAS, MAGIC and Fermi data of the Crab pulsar. Plot is from \cite{OtteSci2011}. Upper limits from other experiments are also shown.  Residuals of both the exponential cut-off and broken power-law fits are also shown.   } 
%\end{wrapfigure}
\end{figure}

\section{SEARCH FOR CORRELATIONS WITH GIANT RADIO PULSES}

Giant Radio Pulses (GRPs) are a phenomenia that has only so far been seen in $\sim$10 pulsars, including the Crab \cite{CrabGRP}. GRPs are partially characterized by possessing a very high flux density relative to average radio pulses. A model by Lyutikov (2007) \cite{Lyutikov2007} explained inter pulse GRPs from the Crab by the presence of elections with Lorentz factors of $10^{8}$ in the outer magnetosphere, which would imply enhanced gamma-ray emission \cite{Lyutikov2007}.

To study the enhanced gamma ray emission corresponding to GRPs, VERITAS took 11.6 hours of simultaneous data with the Green Bank Telescope (GBT), observing 15366 overlapping GRPs. No significant enhancement of the VERITAS pulsar signal was found coincident in this study, so upper limits were calculated from Monte Carlo enchanced Crab signals. UL values for the inter-pulse (IP) and the combined searches are roughly 5 to 10 times the average pulsar flux on single period time-scales. Limits for the main pulse enhancement is much weaker since the number of MP GRPs at 8.8 GHz is $\sim$30 times lower than the number of IP GRPs. Limits are better constrained to 2-3 times the Crab pulsar flux at the widest search windows, but are likely less meaningful theoretically since they nearly encompase the entire simultaneous VERITAS data set \cite{VerGRP}. Models of GRP emission to date provide no quantitative guidance to put limits into a reasonable pulsar/GRP parameter space. 

%\begin{wrapfigure}{L}{0.5\textwidth}
%\begin{figure}
%\usepackage{sidecap}
%\centering
%\includegraphics[width=.5\textwidth]{McCannGRPULs.png}
%\caption{Upper limits of flux enhancement of Crab pulsar emission corresponding to GRP emission in Crab pulsar units vs. the size of the search window used to search for enhancement. }
%\end{wrapfigure}
%\end{figure}

\section{TEST FOR LORENTZ INVARIANCE VIOLATIONS (LIV)}
It has been suggested by several quantum gravity models that due to a possible foamy nature of spacetime, the speed of light in a vacuum could vary with the particle's energy (for a review, see \cite{LIV}). The energy scale for these violations, $E_{n}$, could therefore be constrained by time of arrival differences between photons of different energy orginating from the same source. Testing for these violations requires sources seen at very high energies, with fast variability that are seen at astronomical distances, such as the Crab pulsar above 120 GeV. 

The pulse profiles of the VERITAS data above 120 GeV is then compared to the pulse profile of the Fermi LAT data above 100 MeV. If the same timing solutions are used for both data sets, then the peak positions argee within statistical uncertainty. This indicates no measurable violations of Lorentz Invariance, so a lower limit of the potential LI violations is  therefore calculated. The 95\% confidence upper limit on the timing of the peaks is calculated to be less than 100$\mu$s. The limits of the linear LIV term is therefore:
\begin{equation}
E_{1}> \frac{d \Delta E}{c_{0}  \Delta t_{95\%}} = \frac{2kpc * 120 GeV}{3x10^{8}m s^{-1} * 100\mu s} \sim 3x10^{17} GeV
\end{equation} 
This limit is roughly two orders of magnitude below the Planck mass and the most contraining limit from Fermi using GRBs \cite{FermiGRBLIV}.  The full discussion on the derviation of the result shown here and it's implications with regards to other LIV limits are in \cite{OtteICRC2011}. 

\section{ACKNOWLEDGMENTS}
This research is supported by grants from the U.S. Department of Energy Office of Science, the U.S. National Science Foundation and the Smithsonian Institution, by NSERC in Canada, by Science Foundation Ireland (SFI 10/RFP/AST2748) and by STFC in the U.K. We acknowledge the excellent work of the technical support staff at the Fred Lawrence Whipple Observatory and at the collaborating institutions in the construction and operation of the instrument.

\end{document}